\documentclass[preprint,journal]{vgtc}            

\onlineid{1683}

\vgtccategory{Area 1: Theoretical \& Empirical}

\vgtcpapertype{}

\title{A Multimodal Framework for Understanding\\ Collaborative Design Processes}

\author{%
  \authororcid{Maurice Koch}{0000-0003-0469-8971},
  \authororcid{Nelusa Pathmanathan}{0000-0002-6848-8554},
  \authororcid{Daniel Weiskopf}{0000-0003-1174-1026}, and
  \authororcid{Kuno Kurzhals}{0000-0003-4919-4582}
}

\authorfooter{
    \item 
    All authors are with Visualization Research Center (VISUS), University of Stuttgart. 
    E-mail: \{first name\}.\{last name\}@visus.uni-stuttgart.de.
}

\abstract{
An essential task in analyzing collaborative design processes, such as those that are part of workshops in design studies, is identifying design outcomes and understanding how the collaboration between participants formed the results and led to decision-making.
However, findings are typically restricted to a consolidated textual form based on notes from interviews or observations.
A challenge arises from integrating different sources of observations, leading to large amounts and heterogeneity of collected data.
To address this challenge we propose a practical, modular, and adaptable framework of workshop setup, multimodal data acquisition, AI-based artifact extraction, and visual analysis.
Our interactive visual analysis system, reCAPit, allows the flexible combination of different modalities, including video, audio, notes, or gaze, to analyze and communicate important workshop findings.
A multimodal streamgraph displays activity and attention in the working area, temporally aligned topic cards summarize participants' discussions, and drill-down techniques allow inspecting raw data of included sources.
As part of our research, we conducted six workshops across different themes ranging from social science research on urban planning to a design study on band-practice visualization. 
The latter two are examined in detail and described as case studies.
Further, we present considerations for planning workshops and challenges that we derive from our own experience and the interviews we conducted with workshop experts.
Our research extends existing methodology of collaborative design workshops by promoting data-rich acquisition of multimodal observations, combined AI-based extraction and interactive visual analysis, and transparent dissemination of results.
}


\teaser{
  \centering
  \includegraphics[width=1\linewidth, alt={A view of a city with buildings peeking out of the clouds.}]{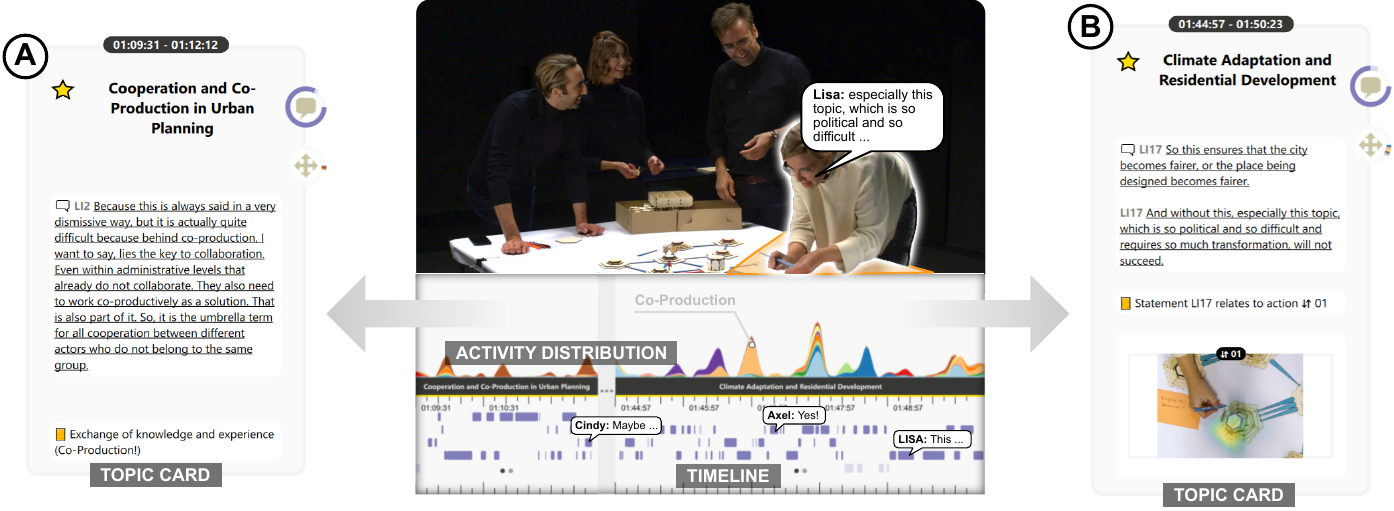}
  \caption{%
  We record multimodal data from collaborative workshops to derive artifacts from discussions. 
  Streamgraphs show activity and visual attention to different regions of the working area. Activity timelines indicate when different roles in the workshop contributed to a discussion. Editable topic cards are derived automatically from transcripts, serving as elements to build topic summaries. 
  The examples show four participants who discuss the topic of \emph{Co-Production} in urban planning. Lisa is currently active in the part of the working area that concerns Co-Production (\textcolor{colorCoProduction}{\CIRCLE}). Topic Cards (A) and (B) provide a multimodal summary of two key events concerning this discussion that the analyst has authored.
  (\textit{Images edited with Stable Diffusion for privacy.})}
  \label{fig:teaser}
}

\graphicspath{{figs/}{figures_cleaned/}{pictures/}{images/}{./}} 

\PassOptionsToPackage{pdftex,dvipsnames,svgnames,table}{xcolor}
\usepackage{xcolor}
\usepackage{multirow}                  
\usepackage{booktabs}                  
\usepackage{amsfonts}
\usepackage{fontawesome5}
\usepackage{makecell}
\usepackage{siunitx}
\usepackage{tikz}
\usepackage{wasysym}
\usepackage{mathptmx}                 
\usepackage{tabularx,ragged2e}
\usepackage{tikz}

\definecolor{colorAudio}{rgb}{0.53,0.67, 0.87}
\definecolor{colorVideo}{rgb}{1.0,0.6, 0.33}
\definecolor{colorGaze}{rgb}{0.82,0.84, 0.0}
\definecolor{colorNotes}{rgb}{1.0,0.67, 0.67}

\definecolor{colorBob}{RGB}{245,120,96}
\definecolor{colorJoe}{RGB}{120,195,176}
\definecolor{colorRandy}{RGB}{230,180,67}
\definecolor{colorCarl}{RGB}{92,167,91}
\definecolor{colorMike}{RGB}{240,144,182}
\definecolor{colorConclusion}{RGB}{74,104,201}
\definecolor{colorCoProduction}{RGB}{246,192,115}

\newcommand*{\priority}[1]{\begin{tikzpicture}[scale=0.075]%
    \draw (0,0) circle (1);
    \fill[fill opacity=1.0,fill=black] (0,0) -- (90:{#1>0?1:0}) arc (90:90-#1*3.6:1) -- cycle;
    \end{tikzpicture}}

\newcolumntype{C}{>{\Centering}X}    
\newcolumntype{L}{>{\RaggedRight}X}  
\newcolumntype{R}{>{\RaggedLeft}X}  

\newcommand\encircle[1]{%
  \tikz[baseline=(X.base)] 
    \node (X) [draw, shape=circle, inner sep=0] {\scriptsize \strut #1};}

\begin{document}


\firstsection{Introduction}

\maketitle

Design processes in the field of visualization, as well as other domains, are often part of a collaborative effort, comprising multiple stages from acquiring knowledge about potential means to solve a problem to the implementation and analysis of a new design~\cite{sedlmair_design_2012}.
One way to address the earlier stages of the design process is conducting creative visualization-opportunities (CVO) workshops, which typically involve domain and visualization expertise to address the problem at hand~\cite{kerzner_framework_2019}. 

During such workshops, participants create fragments (e.g., physical objects, utterances, sketches) as outcomes that foster discussion and potentially support decision-making. These fragments further serve as a means to document the whole process
by textual summarization of findings, images, or objects.
If researchers want to communicate how these results were derived, a temporal analysis of the workshop is required, by someone taking notes throughout or by recording audio and/or video, leading to an increased analysis effort.

Workshop analysis often relies on open coding \cite{cresswell1998} of recorded audio and handwritten notes, which is time-consuming and requires extensive analysis of the material. The high workload associated with established workflows discourages workshop conductors from recording or harnessing material from other sources, such as video or physiological sensors. However, video has the potential of capturing activities that are not accompanied by verbal utterances, such as gestures, activities, and movement. Many aspects of the video can be analyzed automatically with state-of-the-art machine learning (ML) and artificial intelligence (AI) methods, but an appropriate presentation of the results is required for interpretation by humans. Physiological sensors, such as eye trackers, can measure mutual and shared visual attention, which play an important role during communication \cite{jording_social_2018}.
Currently, there is a lack of guidance for workshop conductors on how to properly instrument workshops, collect, consolidate, and analyze resulting multimodal data.
This work presents a multimodal framework and practical guidance for workshop conductors to address these challenges.

Our integrative visualization approach improves the analysis and documentation of design processes from multimodal data collected during collaborative workshops. 
In particular, we want to address the research question: \textit{How can we capture, summarize, and communicate important findings and the way they were derived from a collaborative design process?}
One example is depicted in Figure~\ref{fig:teaser}, from a workshop in social science research on urban planning.
The participants sketched their ideas on a large poster and used crafting materials to design prototypes. 
With our framework, researchers who conducted the workshop can revisit important events from different perspectives of the multimodal data.
We provide a concise overview of all involved data streams and support a drill-down into each individual stream. 
Our multimodal framework embraces \emph{abundance}~\cite{meyer_criteria_2020} by collecting data from a multitude of modalities (Figure~\ref{fig:workflow}).
We argue that consolidating data from multiple sources provides insights from different angles and a rich, multi-faceted view of design processes.
\emph{Transparency} and \emph{plausibility}~\cite{meyer_criteria_2020} are fostered by documenting and communicating the provenance and process with which results were obtained.
Our approach is designed to work with variable combinations of the proposed sources. 
We deliberately designed our framework with flexibility in mind, as using more data sources often increases the complexity of the setup, and thereby might present a challenge to workshop conductors providing the necessary technical infrastructure.

We aim to provide a collection of automatically derived temporal artifacts from the data that can be explored to create summaries of findings. 
The temporal representation of the data is mapped to a multimodal streamgraph and activity timelines as an overview enriched by selected annotations to summarize different workshop phases, important events, notes, and discussion topics.
Our contributions are technical and methodological:

\begin{itemize}[leftmargin=*]
    \item We present a modular framework that consolidates data recorded from design workshops into a comprehensible format to facilitate the analysis and reporting alike.
    \item reCAPit (Reflecting on Collaborative Design Processes) is a publicly available visualization system that combines the extracted data artifacts from video, audio, notes, and eye tracking. It supports a diverse set of analysis tasks that we identified in close collaboration with workshop experts.
    \item We showcase our approach in two case studies: a design study on band-practice visualization and social science research on urban planning. Further, we provide methodological recommendations on how to apply our technical setup in workshops, based on interviews and our two-year experience accompanying six workshops.
\end{itemize}

Our work contributes to bridging the gap between traditionally separate quantitative and qualitative methods by integrating data-intensive and quantitative aspects, thus combining their different strengths. 
We show that rich, multimodal recordings from workshops can be performed, analyzed, and used to gain a more in-depth understanding of collaborative processes in design studies and other design contexts.

\section{Related Work} \label{sec:related}
Our work is related to visualization approaches for multimodal data, with a focus on video data. 
Further, our work is related to the visualization of collaboration.

\subsection{Workshops and Collaborative Processes}

Multimodal learning analytics is a research area that studies learning processes by integrating multimodal data\cite{Blikstein2013multimodal}, often in co-located settings such as healthcare scenarios \cite{Echeverria2019collaboration, martinez-maldonado_data_2020}.
Echeverria et al.~\cite{Echeverria2019collaboration} propose the concept of a \emph{multimodal matrix}, which organizes multimodal observations into segments and phases. Building on their work, we adopt a similar concept to organizing multimodal data into segments. Martinez-Maldonado et al.~\cite{martinez-maldonado_data_2020} explore how visualization and data storytelling can convey multimodal problem insights into teamwork to students and teachers. While their work focuses on simulation-based training, our approach applies to general collaborative design processes.
Meyer and Dykes~\cite{meyer_criteria_2020} introduced six criteria for rigor in visualization design studies: \textit{informed}, \textit{reflexive}, \textit{abundant}, \textit{plausible}, \textit{resonant}, and \emph{transparent}.
These criteria are grounded in the interpretive view that conducting design studies should not be detached from the views and biases of researchers.
As stated by Meyer and Dykes, workshops can support the \textit{abundant} criterion by fostering the collection of ideas and potential solutions. 
However, resulting fragments of such workshops are often gathered as summative collections of notes, posters, transcripts, etc. One issue with such an approach is that only the final results of the process are captured, while the process itself that led to these results is often neglected or even unavailable.

Previous approaches focused on a holistic understanding of the design study process, with the objective of collecting fragments throughout the course of the design study \cite{rogers_insights_2021}. 
While important, the formation process of insights and conclusions drawn from workshops remains elusive when recorded at such low granularity. 
Considering the richness of design workshops, we argue that such studies would benefit from a rigorous reporting methodology that extracts insights from individual CVO workshops~\cite{meyer_criteria_2020} and other sessions.
Thus, we consider our work as an extension and not a replacement of previous research endeavors for rigor in design studies.
Insights and conclusions gathered from design workshops are valuable to the \textit{winnow}, \textit{cast}, and \textit{discover} stages of the nine-stage framework of the design study methodology~\cite{kerzner_framework_2019, sedlmair_design_2012}.

\subsection{Visualization of Collaboration}

Our work is closely related to open-coding approaches~\cite{cresswell1998,isenberg2008}, which can be found in common qualitative research tools like MAXQDA~\cite{MAXQDA2025}, ELAN~\cite{wittenburg2006}, and NVivo~\cite{NVivo2025}. 
These qualitative research tools offer timeline-based visualization based on defined codes, but lack support for visual analysis of multimodal data. 
There are several approaches applicable to examining collaboration using multi-party discourse visualization. MeetingVis~\cite{shi2021} is a visual narrative approach for summarizing meetings. Similarly, ConToVi~\cite{el-assady_contovi_2016} provides topic-space views for analysis of multi-party conversations. Conversation Clock~\cite{bergstrom_conversation_2007} leverages radial visualization for live feedback of group interactions in co-located spaces. Similar to these works, we also use linguistic features to structure conversations, but propose a more comprehensive analysis by incorporating multimodal data. 

\subsection{Multimodal Data Visualization}

Several works investigated systems for multimodal analysis of UX behavior \cite{batch_uxsense_2024, Blascheck2016, soure_coux_2022}, presentation videos~\cite{zeng_emoco_2020, Wu2020multimodal},  e-commerce videos~\cite{tang_videomoderator_2022}, customer service videos~\cite{wong_anchorage_2024}, or public speaking \cite{Huang2024speech}. These works focused only on a single participant in a desktop scenario or presentation settings, whereas we consider the analysis of collaborative design processes, involving multiple individuals. We adopt the strategy of combining AI and interactive visual analysis for data-rich observations~\cite{weiskopf2024bridging}. 

Few approaches focused on long-term multimodal data and videos. 
Kurzhals et al.~\cite{kurzhals2020} visualized features of long-term eye-tracking videos. The authors displayed low-level gaze and video features as spectrograms, combined with filters to extract similar time spans. While we use a similar data basis, we focus on high-level features (e.g., topics) of the videos. Liebers et al.~\cite{liebers_viscomet_2023} proposed a visual analytics approach based on timelines to visualize gaze, activity, and speech events to compare team performances. We adopt a similar visual scheme to depict dialogues and notes within our activity timeline. However, we also provide consolidated information within topic cards and leverage streamgraphs~\cite{burch2013a,byron2008,havre2000} to depict activity and attention over time.

A few approaches incorporated annotations into analysis processes. uxSense~\cite{batch_uxsense_2024} includes {Annotlette} to report transcript snippets and user-defined annotations. ChronoViz~\cite{Fouse2011chrono} combines multimodal data synchronized with digitized notes, which facilitates navigation during analysis. We employ a similar approach with digital notes recorded during workshops. Gaze Stripes~\cite{kurzhals2015} allow analysts to annotate a timeline visualization of videos and gaze data with different elements showing screenshots and statistics. Our approach incorporates similar reporting functionality but also includes information from other modalities. Further, we use elements of storytelling with visualization, which has become an established means for dissemination~\cite{segel2010,boy2015,dasu2023,zhao2023}. Although not multimodal, TimeLineCurator~\cite{Fulda2016time} is a tool that supports authoring of visual timelines from unstructured text. We employ a similar authoring concept but target multimodal data.

\begin{figure}[t]
\centering
    \includegraphics[width=\linewidth]{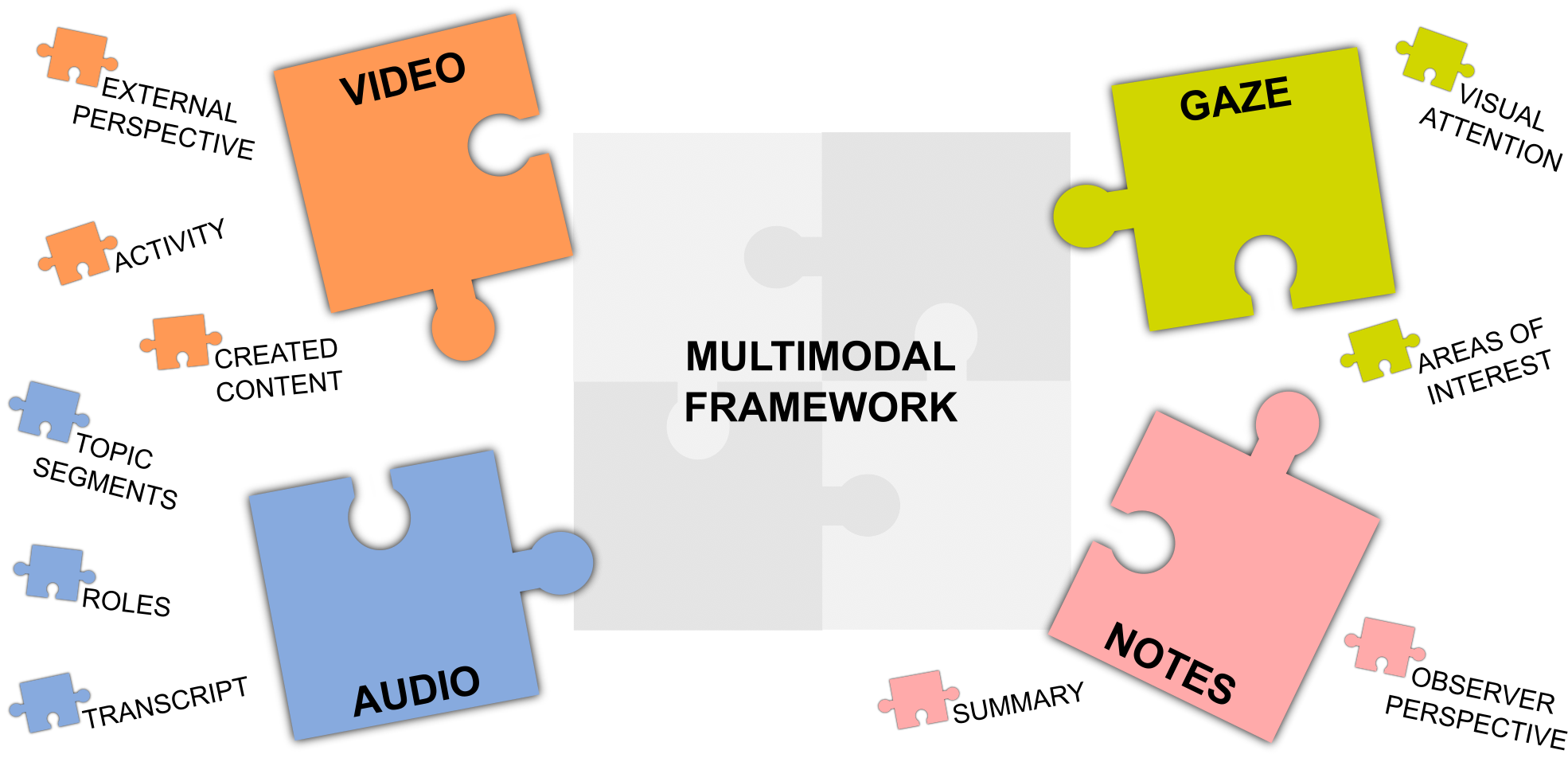}
    \caption{Our framework supports the flexible combination of recording audio, video, notes, and/or gaze for a multimodal view on collaborative workshop processes. From this data, we extract artifacts for~analysis. \vspace{-3ex}}
    \label{fig:workflow}
\end{figure}

\section{Framework Overview}
Our proposed framework extends existing methodology and comprises the recording of data, artifact extraction, visual analysis, and dissemination.
For data acquisition, we decided to include four of the most common sources for behavior analysis: notes, audio, video, and gaze.
Additional sources, e.g., biophysical measurements, would also be possible to integrate but typically require more time-consuming calibration.
From this data, we derived virtual artifacts that help solve different tasks (Figure~\ref{fig:workflow}). These artifacts are of different levels of abstraction from the raw data (e.g., transcripts to topic segments) and, in combination, can serve sensemaking for research questions on a very generalizable level.
Table~\ref{tab:artifacts} summarizes all data sources, derived artifacts, how they are visualized, and the intended task support.
In line with the concepts of \emph{designing as domain experts}~\cite{Fu2024hoop} and \emph{visualization for visualization (Vis4Vis)}~\cite{weiskopf2020Vis4Vis}, our visualization design is informed by our two-year experience that we acquired by accompanying six workshops from different domains and by additional prior experience from own design studies and workshops in visualization research. The visualization is grounded on our expertise in video and eye-tracking visualization with consideration of established analysis tasks and representations for individual artifacts.
We further developed the task support in conjunction with workshop experts from different research fields and through an expert interview, aiming to provide a generalizable approach to address a wide range of research questions. 

\begin{figure}[t]
    \centering
    \includegraphics[width=\linewidth]{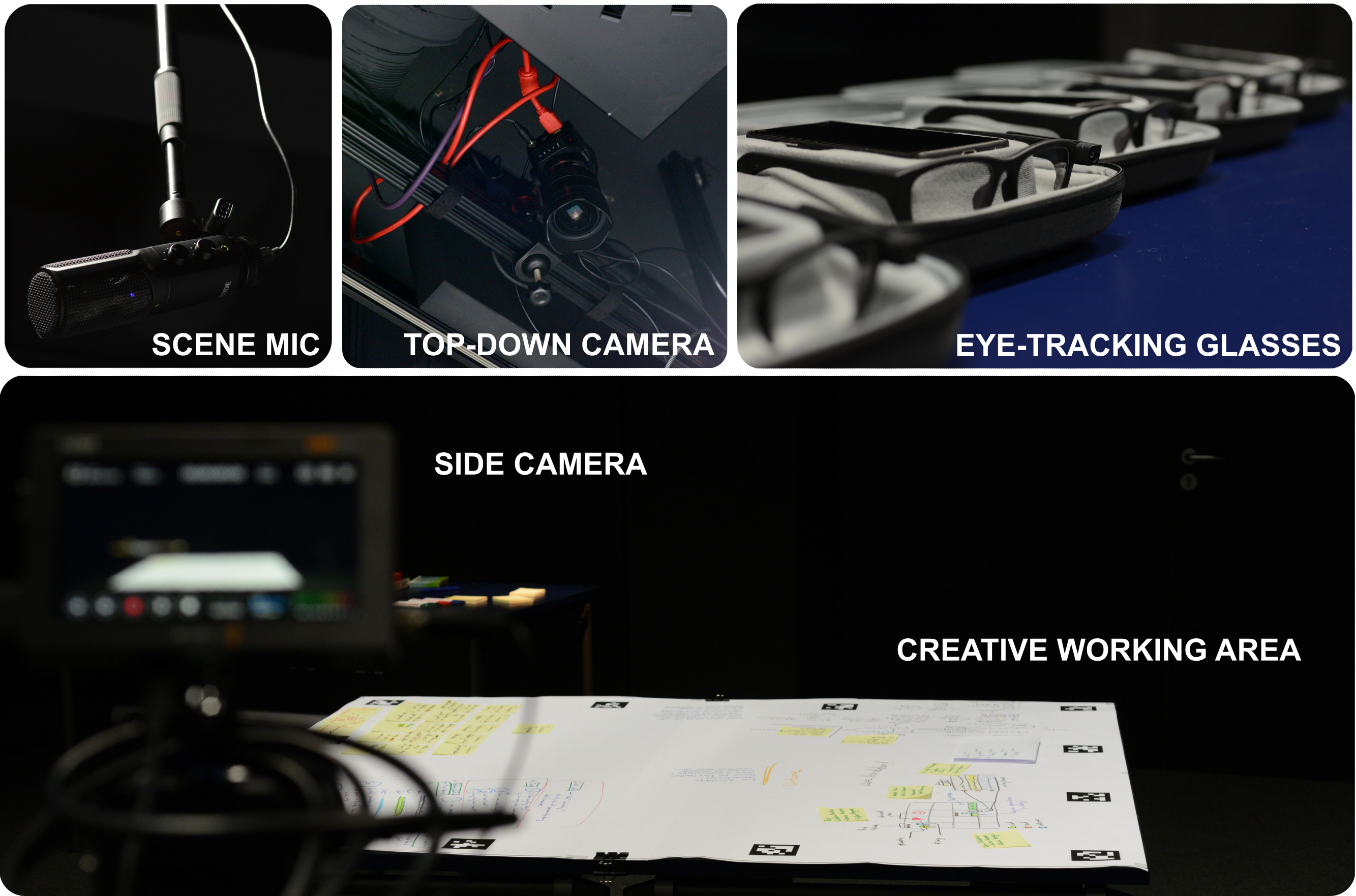}
    \caption{Workshop setup with one side camera and one top-down camera. Audio was recorded with a scene microphone mounted above the work area, gaze with multiple eye-tracking glasses.}
    \label{fig:table_setup}
\end{figure}
\begin{figure}[t]
    \centering
    \includegraphics[width=\linewidth]{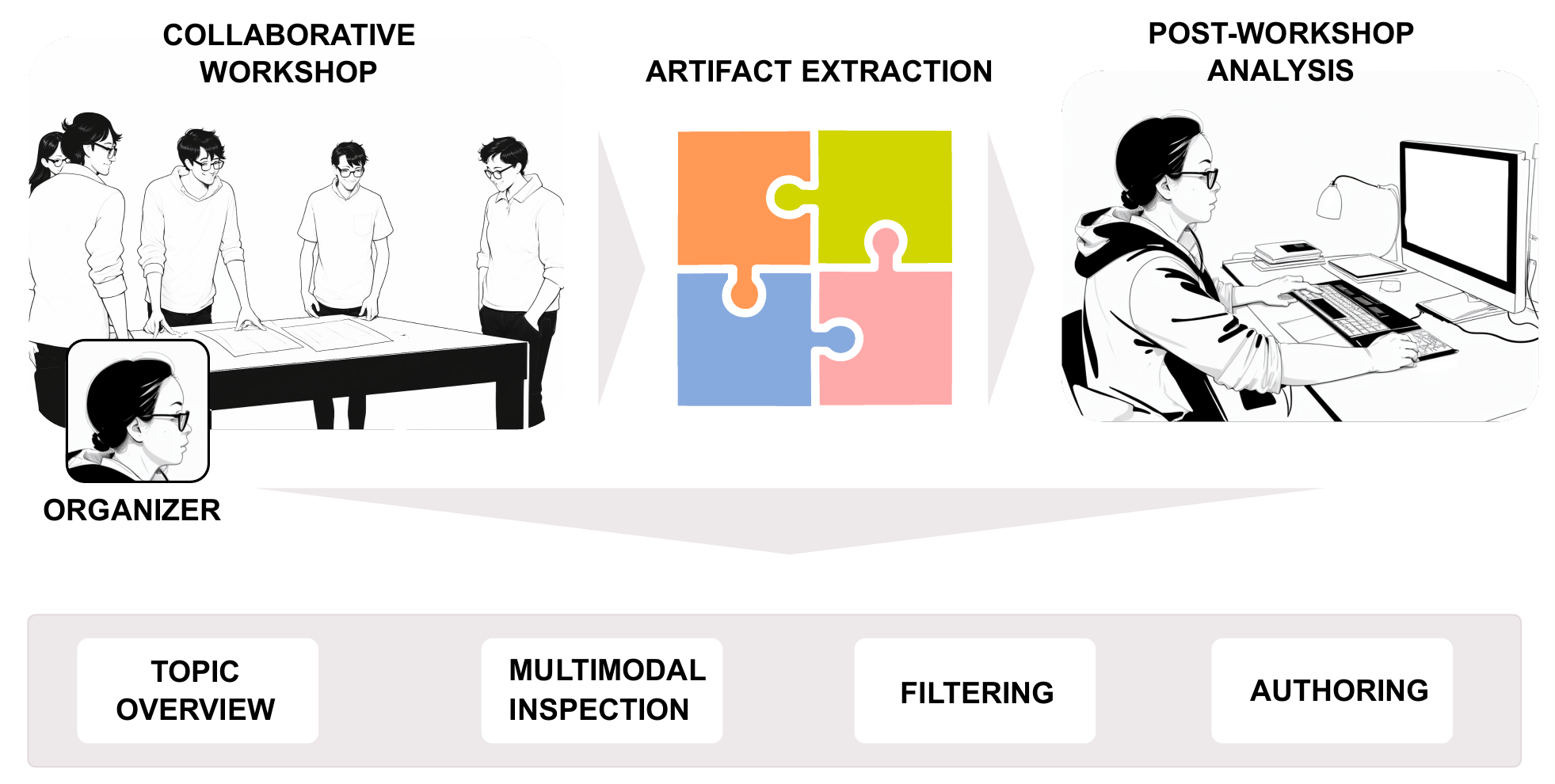}
    \caption{Data is captured during a workshop and processed to extract artifacts. A post-experiment analysis is conducted by workshop organizers. The analysis comprises multiple methods to inspect and author artifacts to derive summaries for sensemaking and dissemination. (\textit{Images edited with Stable Diffusion for privacy.})\vspace{-3ex}}
    \label{fig:analysis_process}
\end{figure}

\subsection{Workshop Setup and Data Acquisition}\label{subsec:data_acquisition}

We focus on a creative mapping approach~\cite{knoll_extending_2020} for conducting workshops. This has the advantage that participants work closely in a shared creative working area that can be observed more easily than, for example, different focus groups.
We defined the working area by two tables covered with paper sheets (Figure~\ref{fig:table_setup}) and printed fiducial markers along the edges as visual features for reliable tracking of the working area.

\paragraph{Video \color{colorVideo}\faVideo}
We optimized for a top-down perspective of the camera for an overview of the working area without occlusions and perspective distortion, and side cameras for observation studies as a source for numerous artifacts related to body language and behavior.

\paragraph{Audio \color{colorAudio}\faMicrophone} We recorded audio as a separate source with a microphone installed above the working area. This source is synchronized with the video material via an audio-visual trigger signal from a clapperboard.

\paragraph{Gaze \color{colorGaze}\faEye} We recorded gaze with eye-tracking glasses~\cite{tonsen2020high}. These consist of a frame attached to a smartphone as a recording unit. A calibration of the system in the working area is necessary for each participant. It can be performed quickly by a one-point calibration procedure in the center of the table.

\vspace{1ex}
The remaining procedure follows common standards: a moderator introduces the topic and research questions and guides participants through different phases of the workshop. We followed general suggestions beginning with ice-breaker questions~\cite{knoll_extending_2020} and introduction rounds. 
Furthermore, we deployed a second moderator during some workshops to take digital notes \mbox{\color{colorNotes}\faStickyNote}. 

\subsection{Analysis Process} \label{subsec:process}

Figure~\ref{fig:analysis_process} summarizes how our proposed framework addresses data from collaborative workshops.
We differentiate three stages: (1) data acquisition from the collaborative workshop, (2) artifact extraction described in Section~\ref{sec:artifact}, and (3) post-workshop analysis.
Our target users are workshop organizers, allowing their personal observations to be integrated into the analysis.

A multimodal overview of all data streams is challenging to achieve as workshops may last for several hours with many thematic transitions in between.
Hence, we devise a segmentation-based approach that accounts for such thematic changes, while still ensuring sufficient abstraction, to facilitate analysts in navigating through the workshop recording (\textbf{Topic Overview}).

This allows researchers to inspect their data from different perspectives and identify important events and discussions (\textbf{Multimodal Inspection}). When workshop organizers analyze their data, they can typically recall some important events of the workshop or topics that emerged during the discussions. Hence, filtering of topic segments (\textbf{Filtering}) can help identify relevant discussions over time that led to a decision during the workshop. All these findings can be captured with topic cards (Section \ref{subsec:topic_cards}) that researchers can author to include relevant data artifacts to support their reasoning (\textbf{Authoring}).

\section{Artifact Extraction} \label{sec:artifact}

\begin{table*}[t]
\tiny
    \centering
        \caption{Our multimodal framework comprises a set of data sources from which artifacts are extracted either automatically or manually. Artifacts are visualized to provide support for a variety of analysis tasks.}
    \label{tab:artifacts}
    \resizebox{\textwidth}{!}{%
    \begin{tabular}{m{0.7cm} m{0.1cm} m{0.1cm} m{0.1cm} m{0.1cm} m{0.1cm} m{0.1cm} m{0.1cm} m{0.1cm} m{1.4cm} m{3.4cm}}
    \toprule
        \textbf{Data} & \multicolumn{8}{c}{\textbf{Artifact}}&  \textbf{Visualization}& \textbf{Task Support}\\
        \cmidrule(lr){2-9} 
           & \rotatebox[origin=b]{90}{\parbox{0.8cm}{\raggedright AOIs}} & \rotatebox[origin=b]{90}{\parbox{0.8cm}{\raggedright Activity}} & \rotatebox[origin=b]{90}{\parbox{0.8cm}{\raggedright Transcript}} & \rotatebox[origin=b]{90}{\parbox{0.8cm}{\raggedright Roles}} & \rotatebox[origin=b]{90}{\parbox{0.8cm}{\raggedright Notes}} & \rotatebox[origin = b]{90}{\parbox{0.7cm}{\raggedright Topic Segments}} & \rotatebox[origin=b]{90}{\parbox{0.7cm}{\raggedright Visual Attention}} & \rotatebox[origin=b]{90}{\parbox{0.7cm}{\raggedright Screen-\\shots}} & & \\ 
         \midrule  
        
           \textbf{Video}\ \color{colorVideo}\faVideo & \priority{100} & \priority{100} & \priority{0} & \priority{0} & \priority{0} & \priority{100} & \priority{0} & \priority{100} & bounding shapes\newline streamgraph\newline heatmap & AOIs to quantify visual attention and activity (\textbf{T1}) \newline behavior analysis (\textbf{T2})\newline identify changes in the work area (\textbf{T3})  \\
        \midrule
       \textbf{Audio}\ \color{colorAudio}\faMicrophone   &  \priority{0} & \priority{0} & \priority{100} & \priority{100} & \priority{0} & \priority{100}& \priority{0} & \priority{0} & text\newline timeline & read event dialogue (\textbf{T4}) \newline  identify temporal distribution of utterances (\textbf{T5})\\
          \midrule
         \textbf{Gaze}\ \color{colorGaze}\faEye & \priority{0} & \priority{0} & \priority{0} & \priority{0} & \priority{0} & \priority{100} & \priority{100} & \priority{100} &  streamgraph\newline heatmap\newline timeline & find important/ignored areas (\textbf{T6})\newline identify shared attention (\textbf{T7})  \\
        \midrule
         \textbf{Notes}\ \color{colorNotes}\faStickyNote & \priority{0} & \priority{0} & \priority{0} & \priority{0} & \priority{100} & \priority{0} & \priority{0} & \priority{0} & text\newline topic card\newline timeline & include personal observations (\textbf{T8})  \newline note taker comparison (\textbf{T9})\\
         \bottomrule
    \end{tabular}
    }\vspace{-2ex}
\end{table*}

Workshops are often open-ended, and the outcome is unclear, especially in their early stages. Collected artifacts have to be analyzed, which can be very time-consuming and error-prone.
For instance, if videos are recorded, an open-coding approach can be applied where analysts encode behavior and events with high effort to later assess the agreement~\cite{banerjee1999b} between different coders. 
We define artifacts as elements generated in a workshop that are used for analytical reasoning.
This comprises different temporal phases of discussion during the workshop and fragments such as notes and sketches by the participants and the workshop moderators. Such artifacts can be derived automatically or manually.
Figure~\ref{fig:workflow} emphasizes the multimodal and modular nature of our framework that supports the integration of video, audio, eye tracking, and digital notes.  
While we advocate recording more, in accordance with \emph{abundance}, our framework provides the flexibility to accommodate the diverse requirements of workshop organizers.

\subsection{Areas of Interest \color{colorVideo}\faVideo} \label{subsec:aois}

The creative working area provides a collaborative space for workshop participants to interact with tangible objects or record their ideas using drawings or post-its.
Workshops often involve hand drawings and use untraditional materials, such as custom-made objects, which are hard to detect using pre-trained object detection and segmentation models. Furthermore, object tracking is often impaired by occlusion caused by participants while interacting with objects in the working area.
To address these issues, we predefine static areas of interest (AOIs) that partition the working area into semantically meaningful regions.
The AOIs serve as a basis to quantify visual attention and activity in the working area, which we describe in the following sections.

\subsection{Activity \color{colorVideo}\faVideo} \label{subsec:activitiy}
\begin{figure}
    \centering
    \begin{minipage}{0.16\textwidth}
        \centering
        \includegraphics[width=\textwidth]{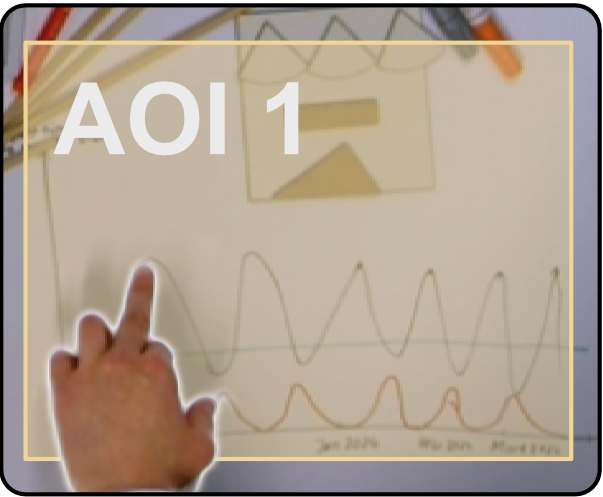}
        \caption*{(a) Pointing}
        \label{fig:hand_pointing}
    \end{minipage}
    \hfill
    \begin{minipage}{0.16\textwidth}
        \centering
        \includegraphics[width=\textwidth]{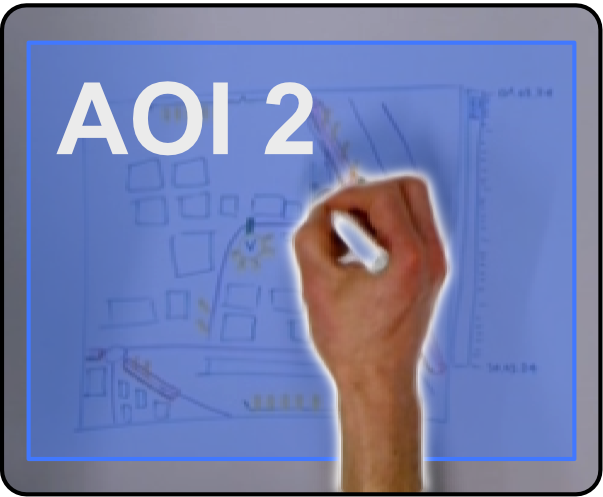}
        \caption*{(b) Drawing}
        \label{fig:hand_drawing}
    \end{minipage}
    \hfill
    \begin{minipage}{0.16\textwidth}
        \centering
        \includegraphics[width=\textwidth]{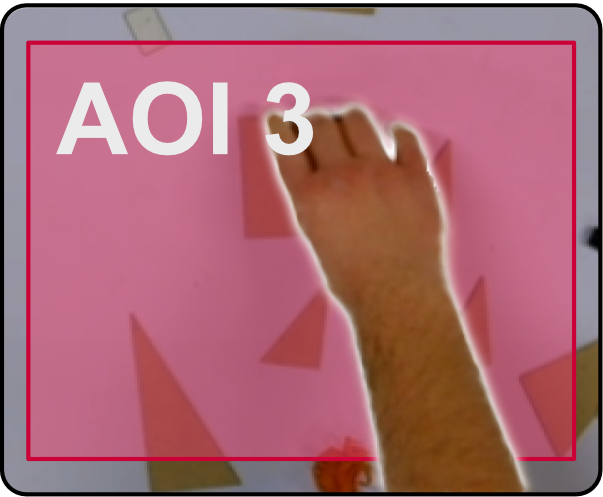}
        \caption*{(c) Interacting}
        \label{fig:hand_interaction}
    \end{minipage}
    \caption{Three common hand gestures are detected in AOIs of the working area to quantify activity. \vspace{-4ex}}
    \label{fig:hand_gesture}
\end{figure}

Participants' interactions within the working area, as captured from the top-down camera, can provide insights into important events, such as pointing or writing gestures shown in Figure~\ref{fig:hand_gesture}. 
It is challenging to robustly and precisely classify different hand gestures, including which participants they originate from, so we focus only on recognizing activity.
Unambiguously identifying participant activity requires either enhanced scene understanding through multiple cameras or assigning color-coded wrist bands to participants.
We argue that despite these limitations, this type of activity recognition can still facilitate detecting phases of individual and joint interactions present in the working area.
We combine background subtraction~\cite{Zivkovic2006adaptive} with hand tracking~\cite{zhang2020mediapipe} to identify activity in the working area.
Background subtraction is an effective method for learning a foreground and background model to separate moving objects (foreground) from still objects (background) in videos. Combined with hand tracking, we can identify activity in the pre-defined AOIs. 

\subsection{Topic Segmentation \color{colorVideo}\faVideo ~\color{colorAudio}\faMicrophone ~\color{colorGaze}\faEye} \label{subsec:segmentation}

\begin{figure}[!t]
\centering
    \includegraphics[width=\linewidth]{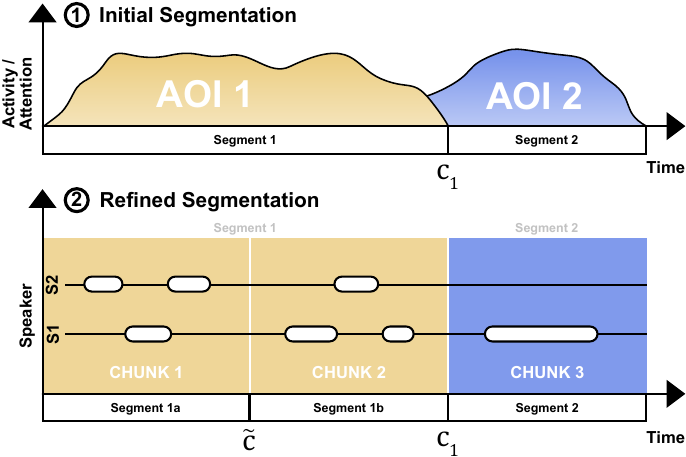}
    \caption{Our multimodal segmentation approach comprises two steps: (1) Initial segmentation based on the activity/attention detected in the working area, and (2) a refined segmentation based on the topic shifts detected in the speaker dialogue. \vspace{-3ex}}
    \label{fig:segmentation-concept}
\end{figure}

Workshops are often organized as full-day events that 
result in several hours of recorded material.
We address this challenge by proposing a two-step multimodal segmentation approach. The approach starts with an initial segmentation, either on the measured activity (Section~\ref{subsec:activitiy}) or visual attention (Section~\ref{subsec:attention}) within the working area. 
The initial segments are refined using detected topic shifts of the dialogues within these segments. Figure~\ref{fig:segmentation-concept} illustrates the two-step segmentation process.

\paragraph{Initial Segmentation \color{colorVideo}\faVideo ~\color{colorGaze}\faEye} We consider a multivariate time series $X = (x_1, \ldots, x_T)$ with $x_i \in \mathbb{R}^M$, where $M$ denotes the number of AOIs and $T$ the number of time steps. This time series represents activity or visual attention.
We apply the Pruned Exact Linear Time (PELT) algorithm~\cite{Killick2012} on the time series to detect change points $C = (c_1, \ldots, c_N)$, where $1 \leq c_i \leq T$. Change points are sorted such that $c_i \leq c_{i+1}$.
The penalty constant $\beta > 0$ controls the number of change points in the PELT algorithm, where higher penalties lead to fewer change points. We empirically found that $\beta = 10$ results in a sufficiently coarse segmentation.

\paragraph{Refined Segmentation \color{colorAudio}\faMicrophone}
The initial segmentation described before can potentially result in a few and very long segments, especially when participants are not interacting in the working area for an extended period.
We address this problem by refining the initial segmentation with detected topic shifts based on the audio transcript.
To this end, we first split the audio transcript within the intervals $[c_i, c_{i+1}]$, where a new chunk is introduced whenever the pause between two consecutive utterances in the audio transcript exceeds a gap threshold.
We use a gap threshold of $1.5$ seconds to account for the fast-paced nature of unstructured group discussions.
Next, we extract text embeddings from chunks with the multilingual GTE model~\cite{zhang2024mgte}. A new change point $\tilde{c}$ with $c_i < \tilde{c} < c_{i+1}$ is introduced when the cosine similarity between consecutive text embeddings drops below a threshold of $0.5$.

\subsection{Visual Attention \color{colorGaze}\faEye} \label{subsec:attention}

Eye tracking provides a means to measure the visual attention of workshop participants.
It offers insights into salient or overlooked areas. 
In particular, joint visual attention is a reliable predictor of collaboration quality \cite{Schneider2018}.
We employ standard eye-tracking methodology and measure visual attention within the working area by detecting fixation hits on the predefined AOIs (Section~\ref{subsec:aois}).

\subsection{Transcript and Roles \color{colorAudio}\faMicrophone}

We use Whisper~\cite{radford2022robustspeechrecognitionlargescale}, a multi-lingual speech recognition model, to transcribe the audio into text.
Since reliable speaker diarization with more than two speakers remains challenging as of today, we performed manual speaker assignments based on the auto-generated transcript. 
Besides speaker assignments, we introduce roles to categorize participants based on professional background, stakeholder role, or seniority. 
Organizing participants in groups has two motivations. First, it reduces the number of encoded colors, which is particularly important with many participants. Second, while individual differences in speaking time between participants can be informative, grouping participants can help analysts identifying patterns emerging at the group level. We further elaborate on the use of participants' roles in Sections~\ref{subsec:timeline} and~\ref{subsec:topic_cards}.

\subsection{Notes \color{colorNotes}\faStickyNote}

Our setup comprises data sources and physiological sensors as an \emph{objective} means to analyze workshops. 
However, workshop conductors often use observations, which is an inherently \emph{subjective} process.
To this end, our framework can incorporate participants' observations as time\-stamped digital notes using an instrumented MS Word document.
Our instrumentation performs versioning of the document by periodically storing copies, allowing for the detection of additions and deletions.

\begin{figure*}[t]
\centering
    \includegraphics[width=\linewidth]{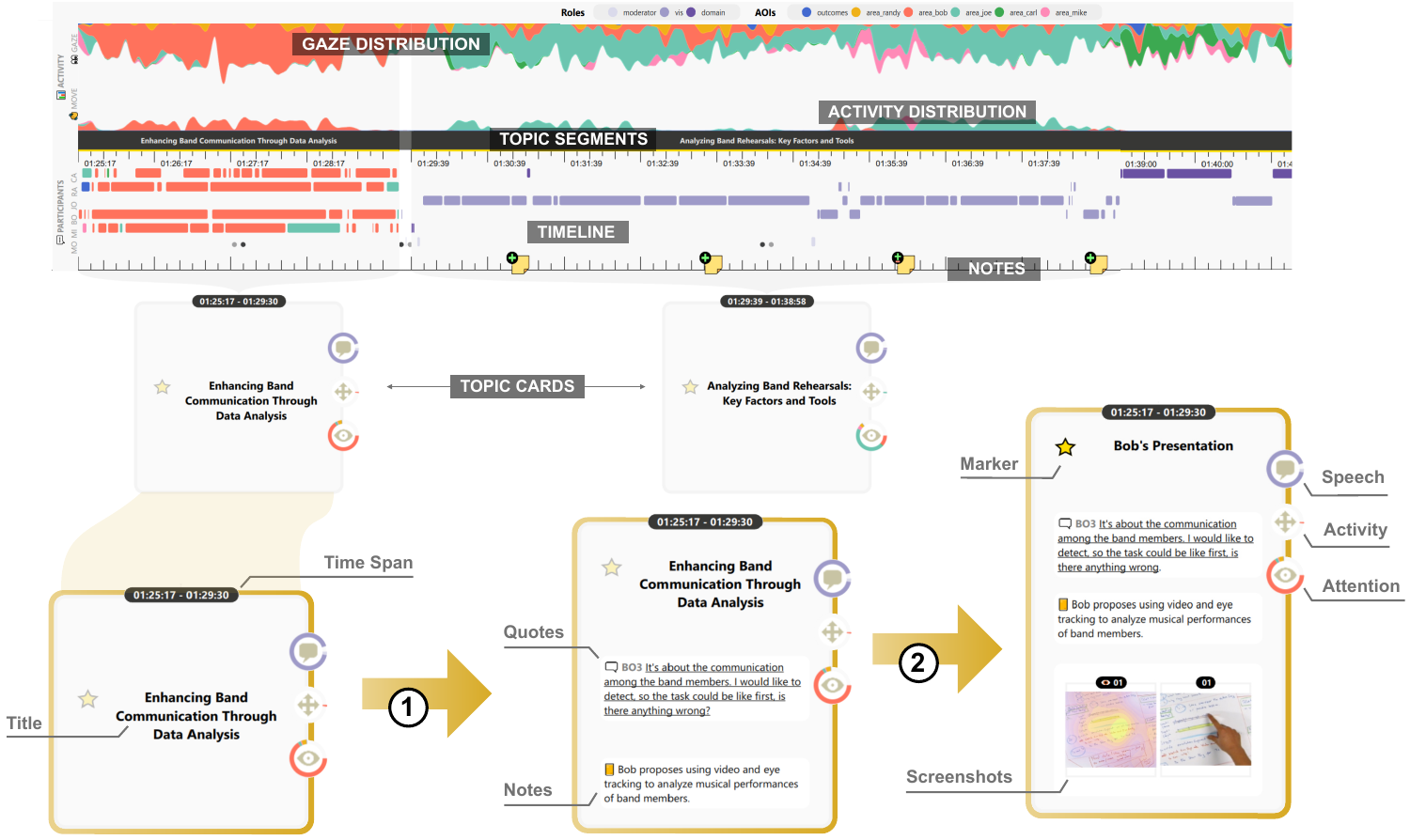}
    \caption{The top depicts our visualization concept on a 15-minute recording comprising gaze, video, audio, and notes data. Two distinct phases (Bob's area, Joe's area) of activity and attention are visible, as indicated by the two detected topic segments, each represented by a topic card. The bottom shows the interactive process of modifying a topic card in step (1) by inserting quotes from the participants' discussion and inserting notes, and in step (2) by defining video crops, adjusting the title, and toggling the marker as a form of bookkeeping. \vspace{-3ex}}
    \label{fig:vis_concept}
\end{figure*}

\section{Visualization System} \label{sec:method}

This work combines the extracted artifacts described in the previous section in a structured overview that can be explored to reflect and analyze the recorded events.
In this section, we present our visual encodings summarized in Table~\ref{tab:artifacts} along with the tasks they support.
Our supplemental material~\cite{koch2025DarusSuppl} contains a more comprehensive discussion on the visualization design.

reCAPit is implemented in Python using Qt 6.6 with PyQt.
The artifact extraction described in the previous section is also implemented in Python, using Sentence Transformers (SBERT)\footnote{\url{https://sbert.net/}}, Media\-Pipe Sol\-utions\footnote{\url{https://ai.google.dev/edge/mediapipe/solutions/guide}} for hand tracking, and OpenCV. 
We used Open\-AI's REST API to create prompts for GPT-4.
Our implementation is modular and supports combining multiple data sources while not relying on specific ones for artifact extraction. Our source code is publicly available at \url{https://github.com/UniStuttgart-VISUS/reCAPit}, with a snapshot of the version used in this paper included in our supplemental material~\cite{koch2025DarusSuppl}.

\subsection{Multimodal Streamgraph}
Previous work \cite{burch2013a} showed the effectiveness of streamgraphs for depicting temporal patterns, particularly for sequential gaze behavior using AOIs.
Similarly, we employ streamgraphs as a temporal overview of how activity and/or attention unfolds (\textbf{T1, T2, T3}). 
Activity and attention are measured within the AOIs. Thus, they serve as the categories of the streamgraphs. 
The value/height in the streamgraph corresponds to the proportion of participants currently fixating on an AOI. AOIs with a peak at 100\% indicate periods of shared attention between all participants. 
For activity, it reflects the percentage of active pixels within an AOI due to hand movements. A peak at 100\% in a given AOI indicates moments when hand movement fully covers the spatial extent of that AOI. 
An example is shown in Figure~\ref{fig:vis_concept}, where there are two distinct phases of attention/activity distributed in Bob's and Joe's areas. 

\subsection{Timeline} \label{subsec:timeline}

The timeline provides task support for multiple sources (Table~\ref{tab:artifacts}).
Topic segments (Section~\ref{subsec:segmentation}) are visually encoded by shaded areas with a top bar displaying the title of a topic segment.
Participants' utterances are represented row-by-row as blocks and color-shaded according to the speaker's assigned role. We chose this visual encoding due to the familiarity of most people with Gantt charts and comparable representations (\textbf{T4, T5}). 
According to Kim et al.~\cite{kim_exploratory_2020}, Gantt charts are one of four main types typically used for conversation visualization, and we use them as they fit well into our timeline-based approach.
Similarly to participants' utterances, we visualize participants' attention row-by-row, which is known in the eye-tracking literature as \emph{scarf-plots}~\cite{richardson2005looking}. This supports the identification of shared visual attention (\textbf{T2, T7}). 
It should be noted that a similar visual encoding is not possible for activity since we only detect the presence of activity but not its source (participant).
The user can interactively switch between utterances and scarfplot views within each topic segment. For example, in Figure~\ref{fig:vis_concept}, the first topic segment displays the participants' scarfplot,  while the second shows participants' utterances.
Timestamped notes recorded during the workshop are placed on the timeline (\textbf{T8}). We use icons to indicate additive \includegraphics[height=1em]{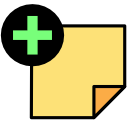} and subtractive \includegraphics[height=1em]{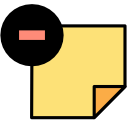} changes to the note document. For changes that involve both addition and subtraction, we use \includegraphics[height=1em]{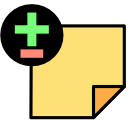}. Multi-note taker comparison can be realized by assigning different colors to note icons (\textbf{T9}).

\subsection{Heatmaps}

Time-based representations are sometimes too detailed, and a more abstract representation is required, especially when finding important/ignored regions in the working area.
To this end, we use \emph{heatmaps} for both activity and attention to complement the multimodal streamgraphs.
Heatmaps are well-known in many domains such as geo-visualization and eye-tracking research. They spatially aggregate attention, often from multiple participants, over a specific period. This facilitates identifying important/ignored areas 
(\textbf{T6}), but can also be insightful behavioral analysis (\textbf{T2}).

\subsection{Topic Cards} \label{subsec:topic_cards}
The extracted topic segments (Section~\ref{subsec:segmentation}) are represented by \textit{topic cards}, showing their temporal order (Figure~\ref{fig:vis_concept}). 
A topic card summarizes the period covered by a segment using multiple elements: title, selected quotes, screenshots, selected notes, and summary statistics of speaker time, activity, and attention.
Most of these elements are customizable, which is an integral part of the authoring process (\textbf{T8}).
We consider topic cards as a means to report the analysis outcomes by contextualizing the visualization with user annotation, borrowing elements from data storytelling~\cite{segel2010} and sharing similarity with previous visualization systems~\cite{batch_uxsense_2024} that integrate similar reporting capabilities.

\paragraph{Title} Titles provide a quick textual summary of the topic segment. Initially, titles are generated by a Large Language Model (LLM) based on the dialog content within the corresponding topic segment. Ideally, the title informs the analyst whether the topic segment is relevant. 

\paragraph{Selected Quotes} 
A topic segment may cover a broad discussion between multiple speakers, but only a subset of the spoken content is interesting. 
The user can add quotes from the utterances associated with the respective topic segment. An utterance identifier is automatically prefixed along with underscores to emphasize verbatim quotes.
\paragraph{Screenshots} As mentioned in Section~\ref{fig:table_setup}, our workshop setup comprises multiple cameras from different perspectives. The user can cut regions out of the video to be displayed in the topic card. Optionally, this screenshot may include an overlay of heatmaps, as shown in the topic card after step \encircle{2} in Figure~\ref{fig:vis_concept}.

\paragraph{Selected Notes} 
Similarly to quotes, the user can put notes into the topic card, which may reference to the selected quotes or any other observations. The user can also access the timestamped notes taken during the workshop, which may also be included.

\paragraph{Summary Statistics} We show how speaking time, activity, and attention are distributed within a topic segment using donut charts. Roles categorize speaker distribution, while AOIs categorize attention and activity. All statistics are normalized by the duration of the segment.

\subsection{Exploration and Analysis}

We now describe how our visualization supports means for analysis as outlined in Section~\ref{subsec:process}: \textbf{Topic Overview}, \textbf{Multimodal Inspection}, \textbf{Filtering}, and \textbf{Authoring}.

\paragraph{Topic Overview} An important part of workshop analysis involves reflecting upon its outcomes and artifacts.
Topic cards provide a visual summary of the workshop data, facilitating an overview of the workshop's structure. Topic cards are visually linked to the timeline.

\paragraph{Multimodal Inspection} Workshops are typically conducted and organized in phases, like individual brainstorming and final idea consolidation. The multimodal streamgraph facilitates identifying such phases since it provides detailed information about the temporal evolution of activity and attention within predefined regions of the working area. Figure~\ref{fig:vis_concept} depicts the streamgraphs during the transition from Bob's working area (\textcolor{colorBob}{\CIRCLE}) to Joe's working area (\textcolor{colorJoe}{\CIRCLE}).

\paragraph{Filtering} 
The number of shown topic cards can impair the analysis. This is particularly the case when analysts are only interested in a subset of discussions. 
We consider keyword search a subsequent step performed after topic overview when the analyst has identified relevant keywords.
To this end, topic cards can be filtered based on user-defined keywords. 
A topic card is displayed when at least one keyword matches the transcript within the topic card's time span. 
The keyword search is case-insensitive and also matches compound words that contain the requested keyword as a substring.

\paragraph{Authoring}

Figure~\ref{fig:vis_concept} depicts the authoring process of the topic segment \emph{Enhancing Band Communication Through Data Analysis}. In step \encircle{1}, the analyst adds a quote from Bob to the topic card along with a note summarizing his contributions. In step \encircle{2}, two screenshots are inserted: one displaying a heatmap overlay that highlights visual attention in Bob's working area, and another capturing the area he points to while presenting his idea. Additionally, the user marks the topic cards, which is a form of bookkeeping. All marked topic segments can be compressed horizontally, as illustrated in Figure~\ref{fig:narrative-segmentation}, effectively filtering out all unmarked topic segments.

\section{Case Studies}\label{sec:case_studies}

Over the last two years, we conducted six workshops across different themes, ranging from social science research on urban planning to a design study on band-practice visualization.
While examining all six workshops is beyond the scope of this paper, we believe that band-practice visualization and urban planning offer complementary insights and unique perspectives into our framework and visualization system. 

\subsection{Workshop on Band-Practice Visualization}

\begin{figure}[t]
    \centering
    \includegraphics[width=.98\linewidth]{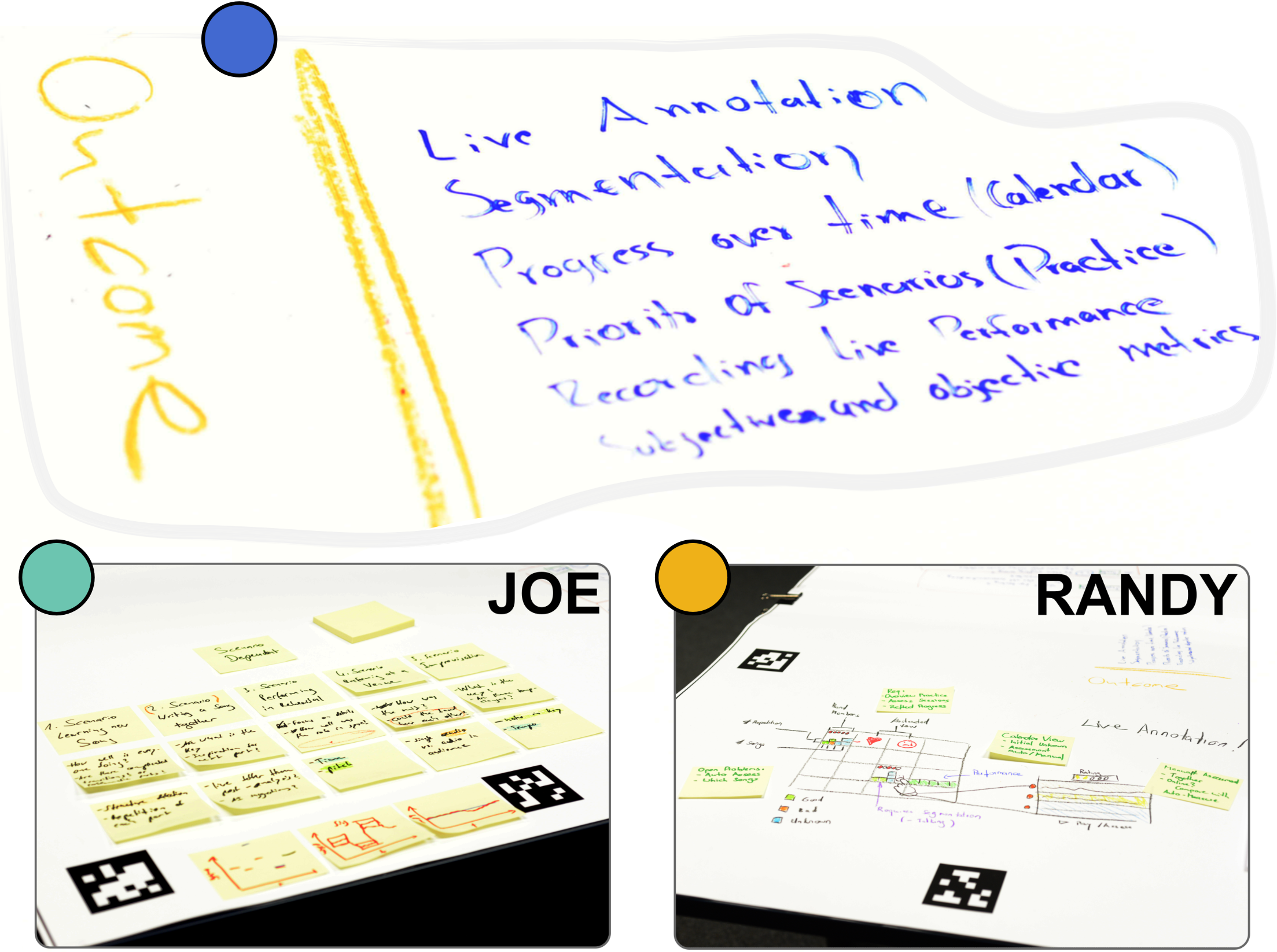}
    \caption{Workshop results of two participants. 
    Notes of different styles were produced, including pure text, sketches, and Post-it notes.
    The outcomes were noted in the center of the working area in the form of a list of requirements. Color-coded circles represent AOIs.\vspace{-3ex}}
    \label{fig:workshop_results}
\end{figure}

\begin{figure*}[h!t] 
\centering
    \includegraphics[width=0.99\linewidth]{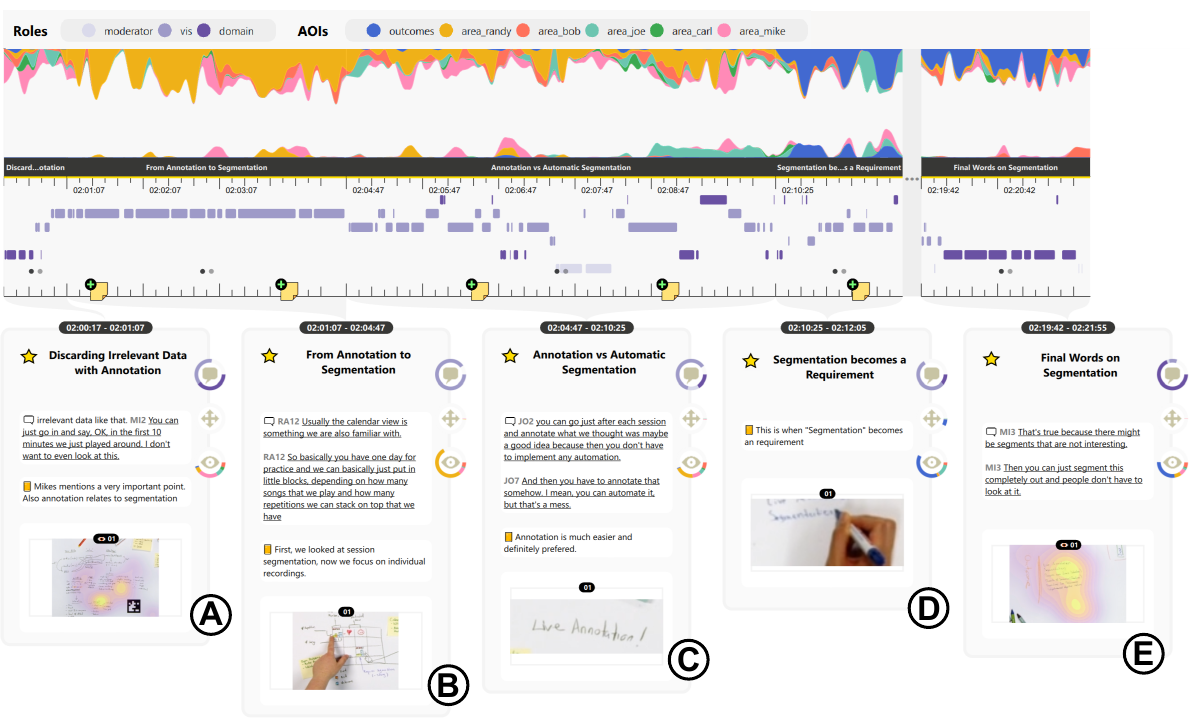}
    \caption{Final analysis state related to the requirement \emph{Segmentation} comprising the five topic segments (A)--(E). The streamgraphs show activity in Randy's area as he explains his visualization and adds the phrase ``Live Annotation!'' to his sketches.\vspace{-3ex}}
    \label{fig:narrative-segmentation}
\end{figure*}

We conducted a workshop at our institute as part of a design study on band-practice visualization. This was the project of another researcher we supported with our framework.
The main research question of this project was: \textit{How can the musicians use the feedback from a visualization to improve their play?} 
The workshop aimed to explore initial ideas for handling data from multiple instruments (guitar, bass, drums, keyboard) and to identify requirements for a future visualization design. 
Within the design study methodology~\cite{sedlmair_design_2012}, this aligns with the \emph{discover} stage.
For this workshop, we recorded video (top-down and side), audio, notes, and gaze.
The individual working areas of the participants served as AOIs (see Figures~\ref{fig:workshop_results} and~\ref{fig:narrative-segmentation}). 

\paragraph{Participants}

The participants were involved in the band-practice project, either as domain experts working with the data or as visualization experts recording the data.
Five participants, all experienced in playing an instrument, attended the workshop. We assigned the following fictitious names to them: \emph{Mike}, \emph{Carl}, \emph{Randy}, \emph{Joe}, and \emph{Bob}. We included one co-author in the workshop to join the discussion actively and lead the conversation during periods of silence. The others were not involved in this paper except for participating in this workshop. 
The goal of the band-practice workshop was to identify possible research directions for Carl, the Master's student supervised by Mike.

\paragraph{Outcomes}

As depicted in Figure~\ref{fig:workshop_results}, the participants used different ways to collect their ideas.
Joe used Post-it notes and started to list different scenarios horizontally, followed by investigating each scenario individually with vertically aligned notes. Randy used a combination of Post-its for specific aspects, such as requirements and problems, and drew sketches of how a visualization could look. 
For a complete overview of all workshop results, we refer to our supplemental material~\cite{koch2025DarusSuppl}.
In this workshop, during the consolidation phase,
the participants summarized the five main requirements for a resulting visualization (see \emph{outcome} in Figure~\ref{fig:workshop_results}).
At this stage, the results are summarized and could be used, for instance, as a list of requirements for a publication. 
How these points were derived is not transparent, except for the final pictures (Figure~\ref{fig:workshop_results}). 
Our approach can enrich the analysis by providing further insight into how the outcome was derived.

\paragraph{Interview (Band Practice)}

We performed Pair Analytics~\cite{hernandez2011} with Carl, who served as subject matter expert.
His task was to guide the authoring process by selecting relevant topic cards, extracting fitting quotes, and adding his subjective perspective using notes.
One author of this paper served as the operator of the visualization tool, while another co-author took notes.
We decided to analyze the requirement \emph{Segmentation [(temporal) of a recording session is necessary]} (second requirement in  \emph{outcome} in Figure~\ref{fig:workshop_results}) with Carl since he expressed particular interest in the segmentation of band-practice recordings.

Before the interview, we introduced Carl to our visualization system and its supported functionalities.
In the following, we outline key aspects of the analysis process, which involves the steps introduced in Section~\ref{subsec:process}. 
The final analysis state is shown in Figure~\ref{fig:narrative-segmentation}.
We wanted to find out how temporal segmentation came up as a requirement.
First, we performed a keyword search (\textbf{Filtering}) based on the terms \textit{segmentation} and \textit{segment}, which yielded the topic segments \encircle{B}, \encircle{D}, and \encircle{E}.
As indicated by the streamgraph, topic segments \encircle{D} and \encircle{E} relate to the consolidation phase in the later stage of the workshop, when the requirements (\textcolor{colorConclusion}{\CIRCLE}) were finalized on the poster (\textbf{Multimodal~Inspection}).
Topic segment \encircle{B} with the original title \textit{Discussing Calendar View and Session Segmentation in Music Practice} caught the interest of Carl.
Segmentation was initially brought up by Randy, who discussed segmentation in the context of multiple sessions.
Randy proposed annotation sessions in the style of a calendar metaphor to track the bands' progress. 
We could verify Randy's participation in this discussion using the streamgraph, which indicated activity and attention in Randy's working area (\textcolor{colorRandy}{\CIRCLE}) when he pointed at his calendar sketch (\textbf{Multimodal Inspection}), which we included as a screenshot (\textbf{Authoring}).
This is when Carl recalled that segmentation and annotation are related concepts, even though they were separated as individual items in the final list of outcomes. 
To this end, we renamed the topic segment \encircle{B} to \textit{From Annotation to Segmentation} (\textbf{Authoring}).
Further, he recollected that Randy's idea of session segmentation later evolved to segmenting individual sessions, which we added as a note.

Keyword search may not identify all topic segments relevant to segmentation.
Thus, the previously identified topic segments served as an anchor point to identify other related topic segments. 
This is when we identified the remaining topic segments \encircle{A} and \encircle{C}, which were more focused on annotation than segmentation (\textbf{Topic Overview}).
In topic segment \encircle{A}, we extracted an interesting statement from Mike: \textit{``[...] removing irrelevant data like that. You can just go and say, OK, in the first 10 minutes we just played around. I don't even look at this.''} Carl agreed that this statement captured the essence of why annotation supports the segmentation of band-practice sessions into digestible parts.
Lastly, in \encircle{C}  we spotted the discussion about automatic segmentation versus annotation, and the general sentiment was in favor of annotation.
We identified in the streamgraph that during this period, \textit{Live annotation} was added to Randy's sketches (\textcolor{colorRandy}{\CIRCLE}).
Carl, an experienced musician himself, also stressed this point, so we added it to the note section accompanied by a screenshot (\textbf{Authoring}).

After we finalized our selection of topic segments, we compressed the timeline to generate the final analysis state shown in Figure~\ref{fig:narrative-segmentation}.
Carl concluded that this summary of the analysis process captures the thought process of segmentation and its relation to live annotation.
From the topic card statistics, we infer that Randy and Joe (both visualization experts) were highly involved in this discussion \encircle{B}--\encircle{D}, while Mike (domain expert) initiated and concluded it.

\subsection{Workshop on Urban Planning} \label{subsec:workshop-urban}

We also accompanied a workshop on urban planning organized by two social scientists of our university, with prior workshop experience. 
The goal of this workshop was to understand the complex relationships between requirements pertaining to urban densification projects.
Due to the multifaceted nature of this project, this workshop included a diverse set of domain experts from politics, the city council, and 
architects. For this workshop, we recorded video (top-down and side) and audio.

We conducted a semi-structured interview with one workshop organizer who had already processed the workshop recordings using his established (coding-based) analysis workflow. 
We refer to Section~\ref{sec:discussion} for a discussion pertaining to the comparison to established analysis workflows. 
In contrast to the previous case study, this one gave us the opportunity to obtain feedback from an expert of a different domain with experience in conducting and analyzing workshops.

\paragraph{Interview (Urban Planning)}

Similar to our previous case study, we presented the workshop recording using our visualization system.
In the subsequent discussion, we refer to two topic cards \encircle{A} and \encircle{B} found in Figure~\ref{fig:teaser}.
We started with identifying discussions concerning the requirement \emph{Co-production}, a concept that actively seeks to include citizens in public service affairs.
The expert was particularly interested in answering the following questions: \emph{How much weight did the workshop participants assign to this requirement, and which results were produced?}
The expert recalled a relevant topic card \encircle{A} when we initially skimmed through the timeline (\textbf{Topic Overview}). He noticed that ``would be really useful'' and ``I would treat that [topic cards] as suggestions.'' In this topic card, he remembered that workshop participant Lisa was explaining the concept of \emph{Co-production} to her peers. He would first get this from the transcript, and then watch the video to better understand exactly what this participant is doing. Then, he stated that it would be important ``that I can record it in notes.'' He emphasizes the importance of modifying topic cards, ``so that you can also capture thoughts and observations''.
Next, our expert spotted another topic card \encircle{B}, where the multimodal streamgraph exhibits an activity peak in \emph{Co-production}, but also substantial activity across other areas, stating there is ``large overlap of these graphs above with the different colors'' (\textbf{Multimodal Inspection}).
The expert remembered from his previous analysis that this was indeed an interesting point during the workshop.
As we looked at the transcript, he noted the problem of \emph{indexicality}, which refers to verbalization when a person is pointing, or describing entities of the scene. An example would be Lisa's statement \emph{``this ensures that the city becomes fairer''}, which is incomprehensible without knowing what \emph{``this''} in the context refers to.
He emphasized the importance of having a video and a heatmap, which facilitates identifying the areas the participants were referring to, stating: ``You can't get that from the audio transcript alone.''
He mentioned that resolving indexicality presents another use case that can be reported using topic cards, something he would usually do within the transcript. In topic card \encircle{B}, we captured the interactions of Lisa, specifically her paintings and drawings concerning the requirement \emph{Co-production} (\textbf{Authoring}).

The expert provided us with positive feedback but also commented on limitations.
First, he noted the limitation of static AOIs, emphasizing the need for continuous object tracking, which ``would have to be optimized a bit for our specific use case.'' Additionally, he requested a feature that allows adjusting the start and end timestamps of topic segments.
Another comment related to a specific research question: whether participants find mutual agreements, particularly in discussions where participants initially held opposing positions. Currently, our framework does not support answering these types of questions, which would require a more in-depth textual analysis of the recorded dialogues.

\section{Discussion}\label{sec:discussion}
Based on our experiences collected during different workshops and the feedback of other workshop organizers, we want to discuss our approach in comparison to alternative methods, how it can be extended in the future, and some practical considerations when planning a workshop based on the proposed framework.

\subsection{Bottom-Up vs Top-Down Analysis}

We identified two main, opposing approaches when analyzing workshops:
(1)~Analysts systematically investigate in a data-driven fashion on a low abstraction level, e.g., with open-coding approaches~\cite{cresswell1998}.
(2)~Analysts mainly investigate the outcomes of a workshop. This can be covered, for instance, by post-workshop questionnaires or pictures of results. If additional data is captured, it is rather investigated to validate hypothesis-driven research or for exploration.
We refer to these opposing directions as \textit{bottom-up} and \textit{top-down} analysis, respectively \cite{pirolli2005,Blascheck2016}. 

\paragraph{Bottom-Up}
Such an approach often depends on the coder's subjective interpretations of things said and done. 
This subjectivity can be put into perspective by measuring multi-coder agreement, at the cost of time-consuming annotation processes.
Our interviewee (Section~\ref{subsec:workshop-urban}) mentioned how communicating and understanding different interpretations can be challenging when discussing the outcomes of the coding process; our visualization aids in conveying individual perspectives on the data, by generating an overview, as well as a structure of key contents from a personal viewpoint.
We further provide means for manual notes and topic summaries that could serve as a basis for codebooks.

\paragraph{Top-Down}
While open coding might take days, our mainly AI-driven data processing can be achieved within 1--2 hours, providing an overview of all data streams. This can significantly lower the initial effort for researchers to include data in their analysis that was formerly omitted only for the infeasibility of the analysis. Our visualization helps recap the events of the workshop without replaying every individual data source. 
As mentioned by our interviewee, the visualization helps find new research questions and answer existing ones efficiently.

Both directions are necessary for a generalizable support of different analysis questions. 
ML models ease the process of artifact extraction, e.g., discussion topics, and reduce the time to derive all important facts. 
However, automatic processing is not always correct, and analysts have to be able to drill down to the raw data to confirm findings and draw their own conclusions. 
Furthermore, reflection about the topics is often already a part of transcription and coding. 
By providing the topic overview, we aim not to remove this reflection from the analysis but rather provide additional perspectives on the data.

\subsection{Alternative Workshop Designs and Setups} \label{subsec:discussion-setup}

Our presented multimodal framework provides workshop conductors the flexibility regarding the data they wish to collect. Hence, we believe that while recording all of the presented data sources might present a challenge to workshop conductors, the complexity of the setup can be scaled accordingly. We constrained the setup to one working area, which might pose limitations to the type of interactions between participants. Further, we do not allow concurrent discussions, which is sometimes found in workshops that precede with a pair-discussion phase. Despite these constraints, the presented setup is only one possible instance, and alternative setups are still covered within our framework. For example, collaboration on wall-mounted whiteboards would only require adjustments to camera positions and angles. Another use case is collaborative sensemaking \cite{mahyar2014support} and Pair Analytics~\cite{hernandez2011} performed either on computer or mobile screens. Additionally, digital pen-based interfaces can replace paper sheets, and as a side-effect, would also simplify detecting handwriting/drawing activity. We also see potential use cases of our approach for analyzing collaborative data analysis \cite{Lam2012empirical}, for example, in immersive environments \cite{ens2021up, cavallo2019immersive}.

\subsection{Additional Data Modalities and Artifacts} \label{subsec:discussion-additional}
Our visualization system can encode a multitude of different signals, and is not limited to the extracted artifacts mentioned in Section~\ref{sec:artifact}. Including additional data sources is achievable with low effort.
This could include biometric measurements, such as skin conductance and heart rate, which could reveal further information about the participants' physical and cognitive states.
Existing data could also be further processed to derive new artifacts.
For instance, we plan to include activity recognition for the video from side cameras~\cite{kong2022}.
The external perspective from the camera could also help delve deeper into aspects of behavior analysis.

\subsection{Considerations for Planning Workshops}
When applying our suggested framework, we want to discuss three main considerations for planning workshops.

\paragraph{Consider Privacy}
The installation of multiple cameras and eye-tracking devices records every movement and gaze of the participants during the experiment. This is highly sensitive data that requires participants' consent and must be handled with care. 
Published material, such as images and videos, usually has to be anonymized. For instance, in Figure~\ref{fig:teaser}, we deployed an image-generative model to anonymize participants~\cite{kurzhals2024}. Additionally, gaze data might also bear personal information that might allow conclusions about the personal traits of the person wearing the glasses~\cite{berkovsky2019}. 

\paragraph{Data Source Selection}

The cost-benefit ratio of including additional modalities in the workshop setup needs to be considered.
This is particularly relevant for physiological sensors like eye tracking or heart rate monitors that not only come at the cost of producing much data (typically from 25 to 100\,Hz), but also can be intrusive, causing participants' discomfort.
We recommend that workshop organizers carefully assess the cost-benefit ratio, but would also like to advocate for recording more data in accordance with \emph{abundance}.
We believe our proposed multimodal framework presents a practical solution to a problem that many workshop organizers face when analyzing their data.
The selection of data sources also depends on the anticipated workshop outcomes and analysis goals (Table~\ref{tab:artifacts}).
Based on our experience and expert interviews, we identified two use cases with complementary objectives: behavioral analysis and dissemination of workshop outcomes.
The latter use case refers to the outcomes \emph{produced} as part of the workshop, often captured by notes, drawings, or other tangible objects. In such cases, installing a top-down camera is recommended.
For behavior analysis, interactions between participants are paramount, which is better captured with side cameras and eye tracking.

\paragraph{Moderation}
For some topics, discussions might become more intense, often resulting in multiple discussions between participants simultaneously. While this is a natural behavior, current algorithms (as well as humans) struggle with transcription from single-source audio recordings in such cases.
This problem could be addressed with individual microphones for each participant, resulting in additional issues with synchronization and potential hardware failures. Hence, for the sake of feasibility and pragmatism, we suggest that moderators should remind participants in such situations to let each person speak without interruption, similar to online meetings.

\section{Conclusion and Future Work}\label{sec:conclusion}
We presented a methodical framework to record, analyze, and report findings of collaborative design workshops with automatically extracted artifacts.
Artifacts might introduce inaccuracies, which can be assessed by our data drill-down approach.
In the future, we plan to incorporate visualization to communicate such uncertainties.
There is a likelihood of missing events that can be compensated by linear skimming through topic cards, which might become time-consuming.
We want to evaluate the time effort in comparison with traditional coding in further experiments.
Topic segments cover a continuous period of the workshop, which means frequently revisited topics might be fragmented in multiple segments that analysts have to connect.
We plan to introduce higher-level abstractions by leveraging topic modeling techniques.
The presented documentation of analysis processes is a first step, and we believe that incorporating further data-storytelling elements can enhance the accessibility of workshop results to non-experts. Additionally, exploring other reporting formats, such as video or even printed booklets, might present promising future directions.
Another application is real-time feedback during workshops, for example, by augmenting the working area with projected content~\cite{dynamicland2025}.
Increasing rigor and transparency of design processes is an ongoing process, and in our opinion, extending the ways to present how qualitative and quantitative findings were derived contributes strongly to achieving this goal. 
With our approach, we build the basis for future extensions on presenting design processes.

\acknowledgments{%
  This work was funded by the Deutsche Forschungsgemeinschaft (DFG, German Research Foundation) -- Project-ID 449742818, Project-ID 251654672 -- TRR 161 (Project B01), and under Germany’s Excellence Strategy -- EXC 2120/1 -- 390831618. We thank Brigitte Schönberger for granting us the opportunity to interview her as part of this research. Figures~\ref{fig:teaser} and \ref{fig:analysis_process} feature photographs containing individuals, which we anonymized with Stable Diffusion to ensure privacy.
}

\bibliographystyle{abbrv-doi-hyperref}
\bibliography{references_cleaned}

\appendix 

\end{document}